\documentclass[aps,pra,10pt,twocolumn,superscriptaddress,floatfix]{revtex4-2}
\usepackage[ascii]{inputenc}
\usepackage{amsmath,amssymb,amsfonts,amsthm}
\usepackage{tikz}
\usepackage{braket}
\usepackage[caption=false]{subfig}
\usepackage[pdftex,bookmarks=false,colorlinks=true,linkcolor=blue,
citecolor=blue,filecolor=black,urlcolor=blue]{hyperref}

\newcommand{\ud}{\mathrm{d}}
\newcommand{\Tr}{\mathrm{Tr}}

\newcommand{\R}{\mathbb{R}}
\newcommand{\C}{\mathbb{C}}

\DeclareMathOperator{\e}{e}

\graphicspath{{figures/}{circuits/}}

\begin{document}

\title{Classical simulation of quantum circuits using a multi-qubit \\ Bloch vector representation of density matrices}

\author{Qunsheng Huang}
\email{keefe.huang@tum.de}
\affiliation{Technical University of Munich, Department of Informatics, Boltzmannstra{\ss}e 3, 85748 Garching, Germany}
\author{Christian B.~Mendl}
\email{christian.mendl@tum.de}
\affiliation{Technical University of Munich, Department of Informatics, Boltzmannstra{\ss}e 3, 85748 Garching, Germany}
\affiliation{Technical University of Munich, Institute for Advanced Study, Lichtenbergstra{\ss}e 2a, 85748 Garching, Germany}

\date{\today}

\begin{abstract}
In the Bloch sphere picture, one finds the coefficients for expanding a single-qubit density operator in terms of the identity and Pauli matrices. A generalization to $n$ qubits via tensor products represents a density operator by a real vector of length $4^n$, conceptually similar to a statevector. Here, we study this approach for the purpose of quantum circuit simulation, including noise processes. The tensor structure leads to computationally efficient algorithms for applying circuit gates and performing few-qubit quantum operations. In view of variational circuit optimization, we study ``backpropagation'' through a quantum circuit and gradient computation based on this representation, and generalize our analysis to the Lindblad equation for modeling the (non-unitary) time evolution of a density operator.
\end{abstract}

\maketitle

\section{Introduction}

Density operators are capable of describing (thermal) quantum ensembles and non-unitary noise processes \cite{NielsenChuang}. In a textbook-type simulation on classical computers, one would store density operators as complex Hermitian matrices in memory, as is currently implemented in widely-used software libraries \cite{Qiskit2019, Cirq, Li2020}. Here, we advocate and study an alternative approach, namely directly working with a tensorized Bloch vector representation, i.e., an expansion in terms of Pauli strings, see Eq.~\eqref{eq:Bloch_nqubit} below. As general insight, this form leads to equations analogous to statevector simulations for quantum circuits, where the multi-qubit Bloch vector assumes the role of the quantum state, and operations on density operators (like applying a unitary matrix by conjugation) become matrix-vector products. The data layout in memory is well suited for single- or two-qubit quantum gates due to the tensor structure, as compared to a literal implementation of matrix conjugations, which involves products from the left and right. As additional advantages, the Bloch representation involves only real-valued quantities, applying general quantum channels does not require a summation over Kraus operators, and gradient computation with respect to gate parameters (see Sect.~\ref{sec:backpropagation}) becomes conceptually simpler.

Generalizations of the Bloch sphere representation for higher-level systems or multiple qubits have been investigated in various forms \cite{Kimura2003, Bertlmann2008, Jevtic2014, Gamel2016}, and the observation that tensor products of Pauli matrices with real coefficients form a basis of Hermitian matrices can be considered common knowledge. Our main contributions here are efficient algorithms and practical details for quantum circuit simulation and variational optimization based on this representation. To clarify, the terms ``density operator'' and ``density matrix'' are used synonymously.

We have implemented the methods described in this work in a Julia software toolbox called \emph{Qaintum} \cite{Qaintum}. As demonstration, we perform parametric optimization of a density matrix via the variational quantum thermalizer (VQT) algorithm described in \cite{Verdon2019}, see Sect.~\ref{sec:vqt_demonstration}.

\section{Tensorized Bloch representation for multiple qubits}

Let us recall the well-known Bloch sphere representation for density matrices: the Bloch vector $\vec{r} \in \R^3$ associated with a single-qubit density matrix $\rho$ is defined via the relation
\begin{equation}
\label{eq:bloch_single_qubit}
\rho = \frac{1}{2} (I_2 + \vec{r} \cdot \vec{\sigma}),
\end{equation}
where $\vec{\sigma} = (X, Y, Z)$ is the Pauli vector and $I_2$ the $2 \times 2$ identity matrix. The property that $\rho$ is positive semidefinite is equivalent to $\lVert\vec{r}\rVert \le 1$ \cite{NielsenChuang}.

By setting $r_0 = 1$ and $\sigma_0 = I_2$, we can rewrite Eq.~\eqref{eq:bloch_single_qubit} as
\begin{equation}
\rho = \frac{1}{2} \sum_{j=0}^3 r_j \sigma_j.
\end{equation}
Slightly more generally, one observes that the vector space of Hermitian $2 \times 2$ matrices is isomorphic to $\R^4$. We can generalize this construction to an arbitrary number of qubits via tensor products of Pauli matrices: any $n$-qubit density matrix $\rho$ has a unique representation as
\begin{equation}
\label{eq:Bloch_nqubit}
\rho = \frac{1}{2^n} \sum_{j_{n-1}=0}^3 \cdots \sum_{j_0=0}^3 r_{j_{n-1},\dots, j_0} \, \sigma_{j_{n-1}} \otimes \cdots \otimes \sigma_{j_0}.
\end{equation}
We will denote the tensor $r \in (\R^4)^{\otimes n} \simeq \R^{4^n}$ as the multi-qubit Bloch vector associated with $\rho$; the condition $\Tr[\rho] = 1$ is then equivalent to $r_{0, \dots, 0} = 1$. For enumerating entries as in $r_{j_{n-1},\dots, j_0}$, we adopt the convention that $j_0$ is the fastest-varying index.

Note that the Pauli strings in Eq.~\eqref{eq:Bloch_nqubit} form an orthonormal basis in the space of Hermitian $2^n \times 2^n$ matrices, with inner product $\langle A \vert B \rangle = \frac{1}{2^n} \Tr[A B]$.

\section{Unitary operations}

A unitary map $U$ acting on a quantum state transforms its density matrix representation by conjugation:
\begin{equation}
\label{eq:rho_U_conjugation}
\rho'= U \rho U^{\dagger}.
\end{equation}
This holds in particular for quantum gates appearing in quantum circuits. Working directly with the Bloch vector representation of $\rho$ in Eq.~\eqref{eq:Bloch_nqubit}, the conjugation~\eqref{eq:rho_U_conjugation} becomes
\begin{equation}
\label{eq:bloch_repr_unitary}
r'= \mathsf{U} r
\end{equation}
when interpreting the Bloch vector $r$ indeed as a vector, $r \in \R^{4^n}$, with $\mathsf{U} \in \R^{4^n \times 4^n}$ an orthogonal matrix. For quantum circuits, which typically involve single- or two-qubit gates, we will provide details for efficient implementations of \eqref{eq:bloch_repr_unitary}, without assembling the matrix $\mathsf{U}$. In the following, the sans-serif styling (as in $\mathsf{U}$) will denote the matrix associated with the Bloch representation in \eqref{eq:bloch_repr_unitary}, given a complex unitary matrix $U \in \C^{2^n \times 2^n}$.

As a concrete example, consider the phase gate $S = \left(\begin{smallmatrix} 1 & 0 \\ 0 & i \end{smallmatrix}\right)$ acting on a single qubit. Since
\begin{equation}
S X S^{\dagger} = Y, \quad S Y S^{\dagger} = -X, \quad S Z S^{\dagger} = Z,
\end{equation}
the conjugation by $S$ in the Bloch representation (including component zero) reads
\begin{equation}
\mathsf{S} =
\begin{pmatrix}
    1 & 0 &  0 & 0 \\
    0 & 0 & -1 & 0 \\
    0 & 1 &  0 & 0 \\
    0 & 0 &  0 & 1
\end{pmatrix}.
\end{equation}
We summarize the corresponding matrix representations related to the Bloch vector formulation for common quantum logic gates in Appendix~\ref{sec:gates_bloch_table}.

\subsection{Application of quantum circuit gates}
\label{sec:circuit_gates}

Let us first consider a single-qubit quantum gate $U^{(1)}$ acting on the $\ell$-th qubit, $\ell \in \{ 0, \dots, n - 1 \}$. The unitary matrix on the full $n$-qubit Hilbert space is thus
\begin{equation}
U = I_{2^{n - 1 - \ell}} \otimes U^{(1)} \otimes I_{2^\ell},
\end{equation}
with $I_m$ denoting the $m \times m$ identity matrix. To efficiently apply this gate to a Bloch vector $r$, we first reshape $r$ into a $4^{n - 1 - \ell} \times 4 \times 4^\ell$ tensor, denoted $\tilde{r}$. Then Eq.~\eqref{eq:bloch_repr_unitary} can be concisely expressed as
\begin{equation}
\label{eq:single_qubit_gate_bloch}
\tilde{r}'_{:, j, :} = \sum_{k=0}^3 \mathsf{U}^{(1)}_{jk} \, \tilde{r}_{:, k, :}, \quad j = 0, \dots, 3,
\end{equation}
where have used the slice index notation ``$:$'' to select all entries along a particular dimension. When specialized for common quantum gates, Eq.~\eqref{eq:single_qubit_gate_bloch} can be implemented in a matrix-free form by expanding the sum and keeping only the non-zero terms of a particular $\mathsf{U}^{(1)}$, cf.\ Appendix~\ref{sec:gates_bloch_table}.

Next, consider a two-qubit gate $U^{(2)}$ acting on qubits $\ell_{\text{a}}$ and $\ell_{\text{b}}$, with $\ell_{\text{a}}, \ell_{\text{b}} \in \{ 0, \dots, n - 1 \}$, $\ell_{\text{a}} < \ell_{\text{b}}$. Now, we reshape $r$ into a $4^{n - 1 - \ell_{\text{b}}} \times 4 \times 4^{\ell_{\text{b}} - \ell_{\text{a}} - 1} \times 4 \times 4^{\ell_{\text{a}}}$ tensor, again denoted $\tilde{r}$. Eq.~\eqref{eq:bloch_repr_unitary} then reads
\begin{equation}
\label{eq:two_qubit_gate_bloch}
\tilde{r}'_{:, j_{\text{b}}, :, j_{\text{a}}, :} = \sum_{k_{\text{a}},k_{\text{b}}=0}^3 \mathsf{U}^{(2)}_{4 j_{\text{b}} +  j_{\text{a}}, 4 k_{\text{b}} +  k_{\text{a}}} \, \tilde{r}_{:, k_{\text{b}}, :, k_{\text{a}}, :}
\end{equation}
for $j_{\text{a}}, j_{\text{b}} = 0, \dots, 3$.

The scheme in Eqs.~\eqref{eq:single_qubit_gate_bloch} and \eqref{eq:two_qubit_gate_bloch} is straightforwardly generalizable to gates acting on a larger number of qubits.

We remark that the matrix-vector form is easier to parallelize as compared to the matrix conjugations in Eq.~\eqref{eq:rho_U_conjugation}, where the multiplications from the left and right would naturally be performed one after another. In terms of memory utilization, the Bloch representation requires the same amount of storage as the upper (or lower) triangular part of a Hermitian matrix, but has a more favorable data layout (for predicting memory access patterns) due to the tensor structure.

A short benchmark comparison is presented in Appendix~\ref{sec:benchmarking}.

\subsection{Controlled gates}
\label{sec:controlled_gates}

Controlled gates turn out to be somewhat tedious to handle when working with density matrices in the Bloch vector representation. Let us first introduce the following ``(anti-)symmetric'' operations acting on the space of Hermitian matrices:
\begin{subequations}
\label{eq:SA_def}
\begin{align}
\mathcal{S}_G(\rho) &= \frac{1}{2}\left( G \rho + \rho G^\dagger \right), \\
\mathcal{A}_G(\rho) &= \frac{i}{2}\left( G \rho - \rho G^\dagger \right),
\end{align}
\end{subequations}
with $G$ a complex matrix of compatible dimension. Bloch representations of $\mathcal{S}_U$, $\mathcal{A}_U$ for common quantum gates $U$ are summarized in Appendix~\ref{sec:gates_bloch_table}.

Now, consider a unitary gate $U$ controlled by a single qubit; this operation can be written as
\begin{equation}
\label{eq:single_ctrl_U}
CU = \ket{0}\bra{0} \otimes I + \ket{1}\bra{1} \otimes U = I + \ket{1}\bra{1} \otimes (U - I).
\end{equation}
For the scenario of $k$ control qubits, $U$ is active only if all of them are in the $\ket{1}$ state (in the computational basis representation). In terms of Eq.~\eqref{eq:single_ctrl_U}, this means generalizing $\ket{1}\bra{1}$ to $\ket{1 \cdots 1}\bra{1 \cdots 1}$ on the right, i.e.,
\begin{equation}
\label{eq:ctrl_U}
CU = I + (\ket{1}\bra{1})^{\otimes k} \otimes (U - I).
\end{equation}
$U$ is always understood to act on the qubits following the control qubits. Application of $CU$ to density matrices leads to
\begin{equation}
\label{eq:ctrl_U_rho}
\begin{split}
&CU \rho \, CU^{\dagger}
 = \rho \\
&\ + (\ket{1}\bra{1})^{\otimes k} \otimes (U - I) \cdot \rho + \rho \cdot (\ket{1}\bra{1})^{\otimes k} \otimes (U^{\dagger} - I) \\
&\ + (\ket{1}\bra{1})^{\otimes k} \otimes (U - I) \cdot \rho \cdot (\ket{1}\bra{1})^{\otimes k} \otimes (U^{\dagger} - I) \\
&= \rho \\
&\ + 2 \mathcal{S}_{(\ket{1}\bra{1})^{\otimes k} \otimes (U - I)}(\rho) \\
&\ + (\ket{1}\bra{1})^{\otimes k} \left(U \rho U^{\dagger} - 2 \mathcal{S}_U(\rho) + \rho \right) (\ket{1}\bra{1})^{\otimes k}.
\end{split}
\end{equation}
Regarding the last term in \eqref{eq:ctrl_U_rho}, note that the conjugations by $\ket{1}\bra{1}$, and evaluation of the expression $U \rho U^{\dagger} - 2 \mathcal{S}_U(\rho) + \rho$, are linear operations acting on different qubits, and thus in particular commute.

In Appendix~\ref{sec:SA_expansion} we verify the following relations (given complex matrices $F$, $G$):
\begin{subequations}
\label{eq:SA_expansion}
\begin{align}
\mathcal{S}_{F \otimes G} &= \mathcal{S}_F \otimes \mathcal{S}_G - \mathcal{A}_F \otimes \mathcal{A}_G, \\
\mathcal{A}_{F \otimes G} &= \mathcal{S}_F \otimes \mathcal{A}_G + \mathcal{A}_F \otimes \mathcal{S}_G.
\end{align}
\end{subequations}
Recursive application allows us to evaluate $\mathcal{S}_{(\ket{1}\bra{1})^{\otimes k} \otimes (U - I)}$ appearing in the penultimate line of Eq.~\eqref{eq:ctrl_U_rho}. Specifically, to expand all combinations of tensor products, we first introduce the shorthand notation
\begin{equation}
\mathcal{E}_G^{(j)} = \begin{cases} \mathcal{S}_G, & j = 0 \\ \mathcal{A}_G, & j = 1 \end{cases}
\end{equation}
Then, based on Eqs.~\eqref{eq:SA_expansion},
\begin{multline}
\label{eq:S_11_U_expansion}
\mathcal{S}_{(\ket{1}\bra{1})^{\otimes k} \otimes (U - I)} \\
= \sum_{j_0, \dots, j_{k-1} = 0}^1 (-1)^{\lceil(j_0 + \dots + j_{k-1}) / 2\rceil} \mathcal{E}_{\ket{1}\bra{1}}^{(j_{k-1})} \otimes \cdots \otimes \mathcal{E}_{\ket{1}\bra{1}}^{(j_0)} \\
\otimes \mathcal{E}_{U - I}^{(j_0 + \dots + j_{k-1} \!\!\!\mod 2)},
\end{multline}
where $\lceil \cdot \rceil$ is the ``ceil'' function (rounding upwards, i.e., closest integer that is greater than or equal to the function argument). In particular, note that the sum in \eqref{eq:S_11_U_expansion} consists of $2^k$ terms, each of which is a tensor product of linear operators acting on separate qubits. Regarding the last operator, also note that $\mathcal{S}_{U - I} = \mathcal{S}_U - \mathrm{id}$ and $\mathcal{A}_{U - I} = \mathcal{A}_U$, which immediately follows from the definitions \eqref{eq:SA_def}.

In summary, we have expanded the conjugation by a controlled gate $CU$, such that operations on the control and target qubits (in the Bloch representation) can be performed sequentially, one control qubit at a time, using the techniques of Sect.~\ref{sec:circuit_gates}. Namely, the conjugations by $\ket{1}\bra{1}$ in the last line of \eqref{eq:ctrl_U_rho} can be applied one by one, and likewise for the penultimate line of \eqref{eq:ctrl_U_rho}: using the expansion in \eqref{eq:S_11_U_expansion}, one can first transform $\rho$ by $\mathcal{E}_{\ket{1}\bra{1}}^{(j_{k-1})}$ (cf.\ the last row of Table~\ref{tab:single_qubit_gates} in the appendix), then the result by $\mathcal{E}_{\ket{1}\bra{1}}^{(j_{k-2})}$ etc.\ up to $\mathcal{E}_{\ket{1}\bra{1}}^{(j_0)}$, and finally by $\mathcal{E}_{U - I}^{(j_0 + \dots + j_{k-1} \!\!\!\mod 2)}$.

\section{Quantum channels and Lindblad equation}

In general, a quantum channel $\mathcal{E}$ acting on a density matrix $\rho$ admits the following Kraus operator representation \cite{NielsenChuang}:
\begin{equation}
\label{eq:quantum_channel_kraus}
\mathcal{E}(\rho) = \sum_k E_k \rho E_k^{\dagger}
\end{equation}
with complex matrices $E_k$, which are denoted Kraus operators. Quantum channels generalize unitary transformations. Since they are likewise linear, we can still represent them in matrix-vector form, analogous to Eq.~\eqref{eq:bloch_repr_unitary}:
\begin{equation}
\label{eq:bloch_repr_channel}
r'= \mathsf{E} r,
\end{equation}
with $r \in \R^{4^n}$ the Bloch vector corresponding to $\rho$, and $\mathsf{E} \in \R^{4^n \times 4^n}$ a real-valued matrix describing the channel; see Table~\ref{tab:single_qubit_channels} for some concrete examples.

For the scenario of a quantum channel affecting one or few qubits within a many-qubit system, observe that the tensor product structure is again preserved by the Bloch representation. In particular, the formulas \eqref{eq:single_qubit_gate_bloch} and \eqref{eq:two_qubit_gate_bloch} for the efficient application are valid for quantum channels as well, after substituting $\mathsf{U}^{(m)}$ by the quantum channel analog $\mathsf{E}^{(m)}$.

An important special case is the time evolution of density matrices when including interactions with the environment, which includes, for example, dissipation. The time dynamics is governed by the following Gorini-Kossakowski-Sudarshan-Lindblad equation \cite{GoriniKossakowskiSudarshan1976, Lindblad1976} (in units of $\hbar = 1$):
\begin{equation}
\label{eq:lindblad}
\frac{\ud}{\ud t} \rho = \mathcal{L}(\rho) = -i [H, \rho] + \sum_q \left( L_q \rho L_q^{\dagger} - \frac{1}{2} \left\{ L_q^{\dagger} L_q, \rho \right\} \right),
\end{equation}
with $[A, B] = A B - B A$, $\{A, B\} = A B + B A$, $H$ the principal system Hamiltonian, and $L_q$ the Lindblad operators. Note that we can use the definitions \eqref{eq:SA_def} to express $i [H, \rho] = 2 \mathcal{A}_H(\rho)$ and $\frac{1}{2} \{ L_q^{\dagger} L_q, \rho \} = \mathcal{S}_{L_q^{\dagger} L_q}(\rho)$.

Let $\mathsf{L}$ be the matrix corresponding to $\mathcal{L}$ in the Bloch representation, such that the Lindblad equation reads
\begin{equation}
\label{eq:lindblad_bloch}
\frac{\ud}{\ud t} r(t) = \mathsf{L} r(t).
\end{equation}
In case $\mathsf{L}$ is time-independent, \eqref{eq:lindblad_bloch} has the formal solution
\begin{equation}
r(t) = \e^{\mathsf{L} t} r(0)
\end{equation}
when starting from some initial state $r(0)$ at $t = 0$. We will revisit the Lindblad equation in the context of gradient computation at the end of the following section.

\section{Backpropagation and gradient computation}
\label{sec:backpropagation}

\begin{figure}[!ht]
\centering
\includegraphics{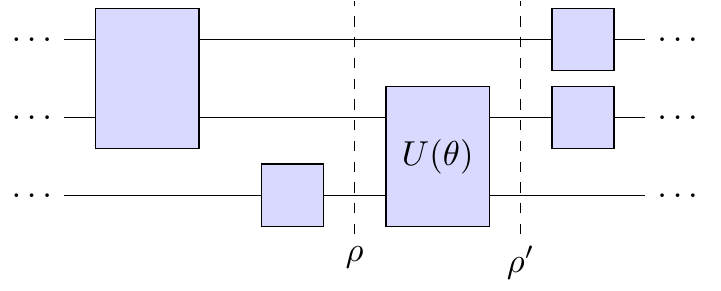}
\caption{Schematic excerpt of a parametrized quantum circuit, with intermediate states $\rho$ and $\rho'$. The blue boxes represent unitary circuit gates.}
\label{fig:circuit_backprop}
\end{figure}

Let $C$ be a real-valued ``cost function'' depending on the output state of a quantum channel $\mathcal{E}_{\text{sys}}$, for example $C = \Tr[M \mathcal{E}_{\text{sys}}(\rho_{\text{in}})]$, with $\rho_{\text{in}}$ the input density matrix and $M$ a measurement operator. For concreteness, we first consider the scenario that $\mathcal{E}_{\text{sys}}$ describes a quantum circuit, such that $\mathcal{E}_{\text{sys}}(\rho) = V \rho V^{\dagger}$, with $V$ the overall unitary transformation effected by the circuit gates -- the general case will be discussed later in this section. Our goal here is to compute the gradient of $C$ with respect to individual parametrized gates in the circuit, which is an essential task for, e.g., variational circuit optimization. For that purpose, we perform a ``backpropagation'' pass through the quantum circuit, which originates from a recursive application of the chain rule for differentiation. Conceptually, in the framework of (classical) artificial neural networks with feedforward architecture, each quantum gate corresponds to a layer in such a network. The setup is sketched in Fig.~\ref{fig:circuit_backprop}, with the density matrix $\rho$ describing an intermediate quantum state, and $\rho' = U(\theta) \rho U(\theta)^{\dagger}$ the next state after applying the parametrized gate $U(\theta)$.

In the following, we use the notation
\begin{equation}
\overline{a} = \frac{\partial C}{\partial a}
\end{equation}
to denote the gradient of $C$ with respect to some variable or parameter $a$ (not to be confused with complex conjugation). We will only encounter real-valued quantities for gradient computation due to the Bloch representation.

Now consider Eq.~\eqref{eq:bloch_repr_unitary}: $r_j' = \sum_k \mathsf{U}_{jk} r_k$ for all $j$. Since $C$ depends on $\mathsf{U}$ only via $r'$, the gradient of $C$ with respect to the entry $\mathsf{U}_{jk}$ obeys
\begin{equation}
\overline{\mathsf{U}_{jk}} = \frac{\partial C}{\partial r_j'} \frac{\partial r_j'}{\partial \mathsf{U}_{jk}} = \overline{r_j'} \, r_k.
\end{equation}
In other words, $\overline{\mathsf{U}}$ is the outer product of $\overline{r'}$ and $r$:
\begin{equation}
\label{eq:grad_U}
\overline{\mathsf{U}} = \overline{r'} \otimes r.
\end{equation}
To obtain the gradient with respect to a $m$-qubit gate $U^{(m)}$, we start from the relation \eqref{eq:single_qubit_gate_bloch}. (The following derivation works analogously for $m \ge 2$; to simplify the notation, we only show the case $m = 1$ here.)
\begin{equation}
\label{eq:grad_Ujk_sum}
\overline{\mathsf{U}^{(1)}_{jk}} = \sum_{u,v} \frac{\partial C}{\partial \tilde{r}'_{u, j, v}} \frac{\partial \tilde{r}'_{u, j, v}}{\partial \mathsf{U}^{(1)}_{jk}} = \sum_{u,v} \overline{\tilde{r}'_{u, j, v}} \, \tilde{r}_{u, k, v}.
\end{equation}
The sum on the right of \eqref{eq:grad_Ujk_sum} can be interpreted as tracing out the remaining qubits (which $U^{(1)}$ leaves invariant). For a general $m$-qubit gate acting on qubits $\ell_{\text{a}}, \ell_{\text{b}}, \dots$, we thus arrive at the formula
\begin{equation}
\label{eq:grad_Um}
\overline{\mathsf{U}^{(m)}} = \Tr_{0:n-1 \backslash \{\ell_{\text{a}}, \ell_{\text{b}}, \dots\}}\big[\overline{r'} \otimes r\big],
\end{equation}
where the partial trace runs over the qubits which are unaffected by $\mathsf{U}^{(m)}$. To efficiently evaluate the partial trace in practice, one can form the entries of $\overline{r'} \otimes r$ ``on the fly'', without storing the outer product as a full matrix.

To complete the gradient computation with respect to unitary gates, let us consider the case that $\mathsf{U}^{(m)}$ depends on real parameters, and denote one such parameter by $\theta$. Then
\begin{equation}
\label{eq:grad_theta_unitary}
\overline{\theta} = \sum_{j,k} \overline{\mathsf{U}^{(m)}_{jk}} \, \frac{\partial}{\partial \theta} \mathsf{U}^{(m)}_{jk} = \Tr\Big[ \overline{\mathsf{U}^{(m)}}^T \frac{\partial}{\partial \theta} \mathsf{U}^{(m)} \Big].
\end{equation}
Specialized to circuit gates, the entries of $\partial \mathsf{U}^{(m)} / \partial \theta$ can usually be evaluated analytically, and one can then efficiently implement the sum in \eqref{eq:grad_theta_unitary} by only keeping the non-zero terms.

A step in the backpropagation requires the computation of $\overline{r}$ based on $\overline{r'}$. Again starting from \eqref{eq:bloch_repr_unitary}, this is achieved by
\begin{equation}
\overline{r_k} = \sum_j \frac{\partial C}{\partial r_j'} \frac{\partial r_j'}{\partial r_k} = \sum_j \overline{r_j'} \, \mathsf{U}_{jk},
\end{equation}
which reads in matrix-vector notation
\begin{equation}
\label{eq:backward_bloch_vector_unitary}
\overline{r} = \mathsf{U}^T \, \overline{r'}.
\end{equation}
Since $\mathsf{U}$ is orthogonal, its transpose is also its inverse, thus \eqref{eq:backward_bloch_vector_unitary} describes the application of the inverse quantum gate to $\overline{r'}$. Directly based on \eqref{eq:bloch_repr_unitary}, the same relation holds for the Bloch vectors as well:
\begin{equation}
r = \mathsf{U}^T \, r'.
\end{equation}
As has been noted before \cite{Luo2020}, one can recompute intermediate quantum states (in our case Bloch vectors) on the fly during the backward pass, which has the potential to significantly decrease computer memory requirements. (For comparison, classical neural networks typically keep the ``activations'' of intermediate layers in memory.) Moreover, we can reuse the techniques in Sect.~\ref{sec:circuit_gates} for the backward pass.

For a general (parametrized) quantum channel $\mathcal{E}$ which maps $\rho' = \mathcal{E}(\rho)$ and is represented by Eq.~\eqref{eq:bloch_repr_channel}, the formulas \eqref{eq:grad_U}, \eqref{eq:grad_Um}, \eqref{eq:grad_theta_unitary}, \eqref{eq:backward_bloch_vector_unitary} literally agree after substituting $\mathsf{U}$ and $\mathsf{U}^{(m)}$ by $\mathsf{E}$ and $\mathsf{E}^{(m)}$, respectively. Namely, the above derivation based on Eq.~\eqref{eq:bloch_repr_unitary} likewise works when starting from Eq.~\eqref{eq:bloch_repr_channel}. However, in the case when $\mathcal{E}$ is not invertible, it is (in general) infeasible to reconstruct $\rho$ from $\rho'$; thus $\rho$ must be kept in memory between the forward and backward pass.

For completeness, we remark that the backpropagation and gradient computation method described here is, in particular, applicable to a composition of quantum channels $\mathcal{E}_{\text{sys}} = \mathcal{E}_1 \circ \mathcal{E}_2 \circ \dots$, again based on the chain rule.

Finally, let us discuss gradient computation based on the Lindblad equation \eqref{eq:lindblad}, which can be regarded as a special case of ``trainable'' differential equations \cite{E2017, Chen2018}. We start from the matrix-vector representation \eqref{eq:lindblad_bloch}, to be solved in the time interval $t \in [0, t_{\text{f}}]$. We assume that the cost function $C$ explicitly depends on the state $r(t_{\text{f}})$ at the final time point. Since $r(t_{\text{f}}) = \e^{\mathsf{L} (t_{\text{f}} - t)} r(t)$ (for time-independent $\mathsf{L}$), it holds that
\begin{equation}
\label{eq:backward_bloch_vector_lindblad}
\overline{r}(t) = \e^{\mathsf{L}^T (t_{\text{f}} - t)} \overline{r}(t_{\text{f}}),
\end{equation}
analogous to \eqref{eq:backward_bloch_vector_unitary}. Thus, $\overline{r}$ obeys the differential equation
\begin{equation}
\label{eq:lindblad_adjoint_diffeq}
\frac{\ud}{\ud t} \overline{r}(t) = -\mathsf{L}^T \, \overline{r}(t),
\end{equation}
which has to be solved backwards in time, with ``initial condition'' $\overline{r}(t_{\text{f}}) = \partial C / \partial r(t_{\text{f}})$. It turns out that \eqref{eq:lindblad_adjoint_diffeq} remains valid for time-dependent $\mathsf{L}$ as well. Namely, one can express \eqref{eq:lindblad_bloch} as ordinary differential equation
\begin{equation}
\frac{\ud}{\ud t} r(t) = f(r(t), t)
\end{equation}
with $f(r, t) = \mathsf{L}(t) \, r$, and then use that the ``adjoint'' $\overline{r}$ is governed by \cite{Pontryagin1962, Chen2018}
\begin{equation}
\frac{\ud}{\ud t} \overline{r}(t)^T = -\overline{r}(t)^T \frac{\partial f(r(t), t)}{\partial r(t)}.
\end{equation}

To relate \eqref{eq:lindblad_adjoint_diffeq} to the original Lindblad equation \eqref{eq:lindblad}, let us define the dual $\mathcal{L}^*$ (acting on Hermitian matrices) via the condition
\begin{equation}
\Tr[\tau \mathcal{L}(\rho)] = \Tr[\mathcal{L}^*(\tau) \rho]
\end{equation}
for all $\tau$, $\rho$. Note that $\mathcal{L}^*$ describes time evolution in the Heisenberg picture and is the analogue of $\mathsf{L}^T$, and thus Eq.~\eqref{eq:lindblad_adjoint_diffeq} can be expressed as
\begin{equation}
\frac{\ud}{\ud t} \overline{\rho} = -\mathcal{L}^*(\overline{\rho}) = -i [H, \overline{\rho}] - \sum_q \left( L_q^{\dagger} \overline{\rho} L_q - \frac{1}{2} \left\{ L_q^{\dagger} L_q, \overline{\rho} \right\} \right).
\end{equation}

To compute the gradient with respect to a time-independent $\mathsf{L}$, we assume that $\mathsf{L}$ is parametrized by some variable $\theta \in \R$, and use the identity \cite{Wilcox1967}
\begin{equation}
\frac{\partial}{\partial \theta} \e^{\mathsf{L} t} = \int_0^t \e^{\mathsf{L} (t - t')} \frac{\partial \mathsf{L}}{\partial \theta} \e^{\mathsf{L} t'} \ud t'.
\end{equation}
Then, by varying a single matrix entry,
\begin{equation}
\label{eq:grad_Ljk}
\begin{split}
\overline{\mathsf{L}_{jk}}
&= \sum_{\ell} \frac{\partial C}{\partial r_\ell(t_{\text{f}})} \frac{\partial r_\ell(t_{\text{f}})}{\partial \mathsf{L}_{jk}} = \overline{r}(t_{\text{f}})^T \frac{\partial}{\partial \mathsf{L}_{jk}} \e^{\mathsf{L} t_{\text{f}}} r(0) \\
&= \overline{r}(t_{\text{f}})^T \int_0^{t_{\text{f}}} \e^{\mathsf{L} (t_{\text{f}} - t)} (e_j \otimes e_k) \e^{\mathsf{L} t} \ud t \, r(0) \\
&= \int_0^{t_{\text{f}}} \overline{r_j}(t) r_k(t) \, \ud t.
\end{split}
\end{equation}
Here $e_j \otimes e_k$ is the matrix with a single non-zero entry $1$ at index $(j,k)$, and we have used the relation \eqref{eq:backward_bloch_vector_lindblad}. Thus, writing \eqref{eq:grad_Ljk} in matrix notation,
\begin{equation}
\overline{\mathsf{L}} = \int_0^{t_{\text{f}}} \overline{r}(t) \otimes r(t) \, \ud t,
\end{equation}
which formally resembles Eq.~\eqref{eq:grad_U}.

Finally, let us discuss gradient computation in the scenario of a time-dependent $\mathsf{L}$. In this setup, $\mathsf{L}$ additionally depends on some parameter $\theta \in \R$, and our goal is computing the gradient of $C$ with respect to $\theta$. We express the Lindblad equation \eqref{eq:lindblad_bloch} as
\begin{equation}
\frac{\ud}{\ud t} r(t) = f(r(t), t, \theta)
\end{equation}
with $f(r, t, \theta) = \mathsf{L}(t, \theta) \, r$. Then, based on the derivation in \cite{Chen2018}, one obtains the following generalization of \eqref{eq:grad_Ljk}:
\begin{equation}
\overline{\theta} = \int_0^{t_{\text{f}}} \overline{r}(t)^T \frac{\partial f(r(t), t, \theta)}{\partial \theta} \,\ud t = \int_0^{t_{\text{f}}} \overline{r}(t)^T \frac{\partial \mathsf{L}(t, \theta)}{\partial \theta} r(t) \,\ud t.
\end{equation}
Expressed in terms of $\mathcal{L}$, this equation reads
\begin{equation}
\label{eq:grad_theta_lindblad}
\overline{\theta} = \int_0^{t_{\text{f}}} \Tr\Big[\overline{\rho}(t) \frac{\partial}{\partial \theta} \mathcal{L}(\rho(t), t, \theta) \Big] \ud t,
\end{equation}
where we follow the convention that $\overline{r}(t)$ is related to $\overline{\rho}(t)$ as in \eqref{eq:Bloch_nqubit} but without the $2^{-n}$ prefactor. Note that $\overline{\rho}(t)$ could be interpreted as measurement operator in Eq.~\eqref{eq:grad_theta_lindblad}

\section{VQT application example}
\label{sec:vqt_demonstration}

In order to demonstrate the practical feasibility of our framework, we implement the variational quantum thermalizer (VQT) algorithm for a quantum Hamiltonian-based model (QHBM) \cite{Verdon2019} using the Qaintum software library. The code for the present example is available at \cite{Code}. The goal is to approximate a target thermal state
\begin{equation}
\label{eq:thermal_sigma_beta}
\sigma_\beta = \frac{1}{\mathcal{Z}_\beta} \e^{-\beta H}, \quad \mathcal{Z}_\beta = \Tr\!\left[ \e^{-\beta H} \right],
\end{equation}
given a known Hamiltonian $H$ and inverse temperature $\beta$.

\begin{figure}[!ht]
\centering
\includegraphics{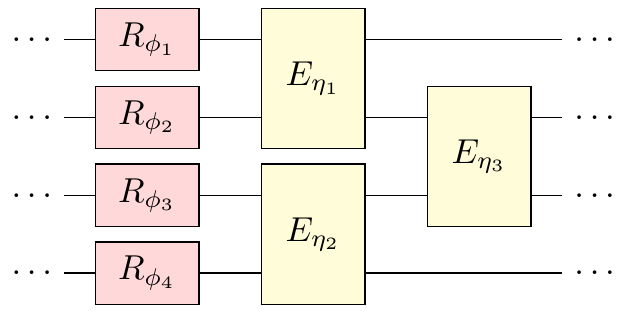}
\caption{A single parametrized layer of the circuit model.}
\label{fig:paramlayer}
\end{figure}

The ansatz density matrix starts from a ``latent'' diagonal density matrix $\rho_\theta$ (parametrized by a real vector $\theta$), which is then conjugated by a unitary matrix $U_{\phi}$ (represented as a quantum circuit with parameters $\phi$) \cite{Verdon2019}:
\begin{equation}
\label{eq:rho_ansatz_def}
\rho_{\theta, \phi} = U_{\phi} \, \rho_{\theta} \, U_{\phi}^\dagger.
\end{equation}
As in the prior work \cite{Verdon2019}, we use
\begin{equation}
\label{eq:latentdensity}
\rho_{\theta} = \bigotimes_{j=1}^n \begin{pmatrix} \frac{1}{2} (1 + \cos(\theta_j)) & 0 \\ 0 & \frac{1}{2} (1 - \cos(\theta_j)) \end{pmatrix}
\end{equation}
for the latent density matrix. Regarding $U_\phi$, the parameterized quantum circuit is a composition of several layers. Each layer in turn consists of two types of gates: a parametrized single-qubit rotation gate on site $j$ defined as
\begin{equation}
R_{\phi_j} = \e^{i (\phi_j^1 X_j + \phi_j^2 Y_j + \phi_j^3 Z_j)},
\end{equation}
and a two-qubit entanglement gate
\begin{equation}
E_{\eta_j} = \e^{i (\eta_j^1 X_j X_{j+1} + \eta_j^2 Y_j Y_{j+1} + \eta_j^3 Z_j Z_{j+1})}.
\end{equation}
As in \cite{Verdon2019}, we use a sequential arrangement of qubits in the circuit with open boundary conditions, independent of the physical model. These entanglement gates are applied in a brick wall pattern. Fig.~\ref{fig:paramlayer} shows a single such layer. For our experiments, we use three layers.

The to-be minimized cost function of the optimization problem is based on the Kullback-Leibler divergence (relative entropy) \cite{Kullback1951} of the ansatz density matrix and the ground truth:
\begin{equation}
\begin{split}
\mathcal{D}(\rho_{\theta, \phi} \parallel \sigma_\beta)
&\equiv - \Tr[\rho_{\theta , \phi} \log(\rho_{\theta ,\phi})] - \Tr[\rho_{\theta , \phi} \log(\sigma_\beta)] \\
&= -S(\rho_{\theta,\phi}) + \beta\,\Tr[H \rho_{\theta,\phi}] + \log(\mathcal{Z}_\beta),
\end{split}
\end{equation}
where $S$ is the von Neumann entropy. Since $\beta$ is fixed, the term $\log(Z_\beta)$ can be regarded as constant for the optimization with respect to $\rho_{\theta, \phi}$. One then arrives at the following cost function:
\begin{equation}
\mathcal{L}_{\theta, \phi} = - S(\rho_{\theta, \phi}) + \beta\,\Tr[H\rho_{\theta , \phi}].
\end{equation}

\begin{figure}[!ht]
\centering
\includegraphics[scale=0.4]{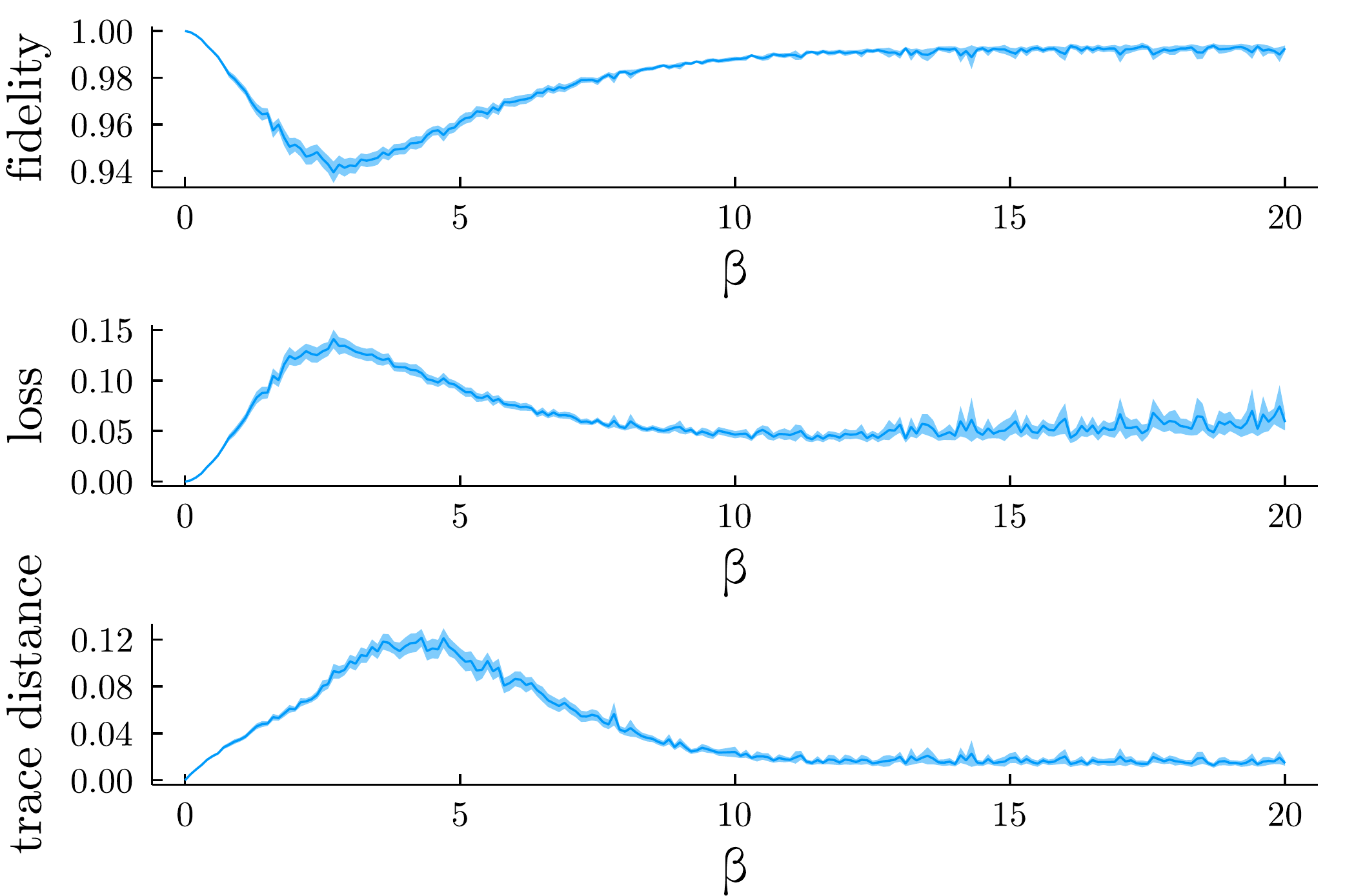}
\caption{Variation of (a) fidelity, (b) loss, and (c) trace distance after 500 optimization iterations over $\beta$ values from $\beta=0$ to $\beta=20$ at $0.1$ intervals for $H_{\text{1D}}$ in \eqref{eq:heisenberg1D} with $n = 4$, with chosen parameters $J=-1$, $g = 0.3$, $h = 0.2$. The shaded region indicates the $95\%$ confidence interval.}
\label{fig:1d_heis_beta}
\end{figure}

For the first experiment, we consider a Heisenberg Hamiltonian on a one-dimensional lattice:
\begin{equation}
\label{eq:heisenberg1D}
H_{\text{1D}} = -J \sum_{j=1}^{n-1} \vec{S}_j \cdot \vec{S}_{j+1} + \sum_{j=1}^n \left( g S^x_j + h S^z_j \right),
\end{equation}
with $\vec{S} = \frac{1}{2} \vec{\sigma}$ on a 1D lattice with 4 qubits ($n = 4$). We kept the $J$, $g$ and $h$ parameters constant and examined how well the model performed when the temperature $\beta$ is varied from $\beta = 0$ to $\beta = 20$ in $0.1$ intervals. We utilized an AdaMax optimzer with learning rate $0.005$ for faster convergence and ran the experiment 50 times with randomized initial parametric values $(\theta, \phi)$ for each $\beta$. The results are shown in Fig.~\ref{fig:1d_heis_beta}. To ensure convergence, we ran each optimization for 500 iterations.

We reproduce the general behavior observed in the previous work \cite{Verdon2019} (a dip in the fidelity and spike in the loss between $\beta = 1$ and $\beta = 5$). As minor remark, in the worst case the minimum fidelity is $\approx 0.93$ here, which is slightly higher than in the prior work. This may be explained by the larger number of optimization steps used here, or a differing gradient-based optimizer.

As next experiment, we apply the optimization procedure to a Heisenberg-type Hamiltonian on a two-dimensional lattice:
\begin{equation}
\label{eq:heisenberg2D}
H_{\text{2D}} = \sum_{\langle j, k \rangle_h} J_h \vec{S}_j \cdot \vec{S}_k + \sum_{\langle j, k \rangle_v} J_v \vec{S}_j \cdot \vec{S}_k,
\end{equation}
where $\langle \cdot, \cdot \rangle_h$ and $\langle \cdot, \cdot \rangle_v$ indicate nearest neighbor pairs in the horizontal and vertical directions, respectively.

\begin{figure}[!ht]
\centering
\includegraphics[width=\columnwidth]{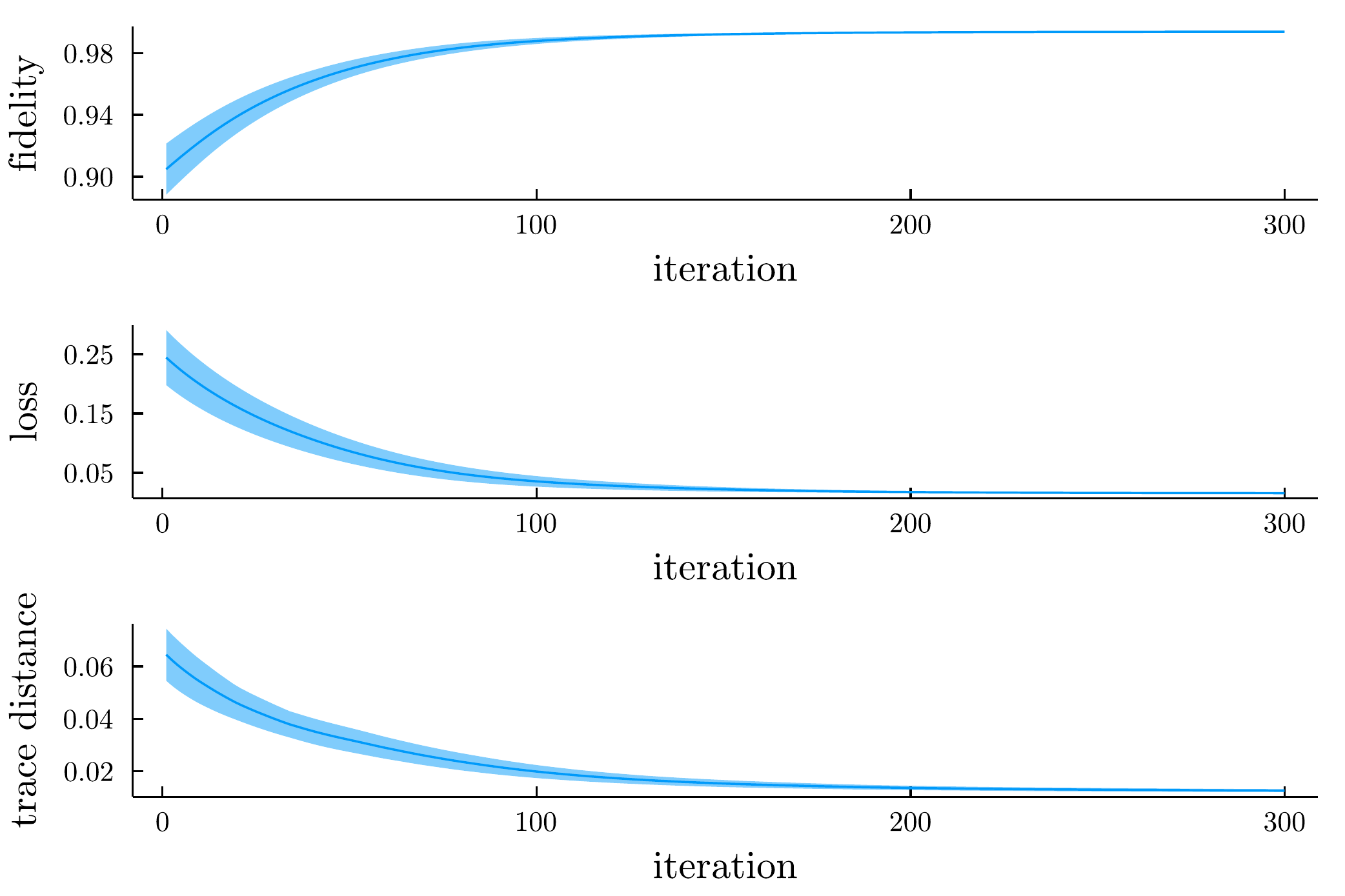}
\caption{Convergence of the VQT algorithm optimization procedure at $\beta=0.5$ for the Heisenberg Hamiltonian \eqref{eq:heisenberg2D} on a $2 \times 2$ lattice with parameters $J_h = 1$, $J_v = 0.6$. The shaded regions indicate $95\%$ confidence intervals.}
\label{fig:2d_heis_iterations}
\end{figure}
\begin{figure}[!ht]
\centering
\includegraphics[width=\columnwidth]{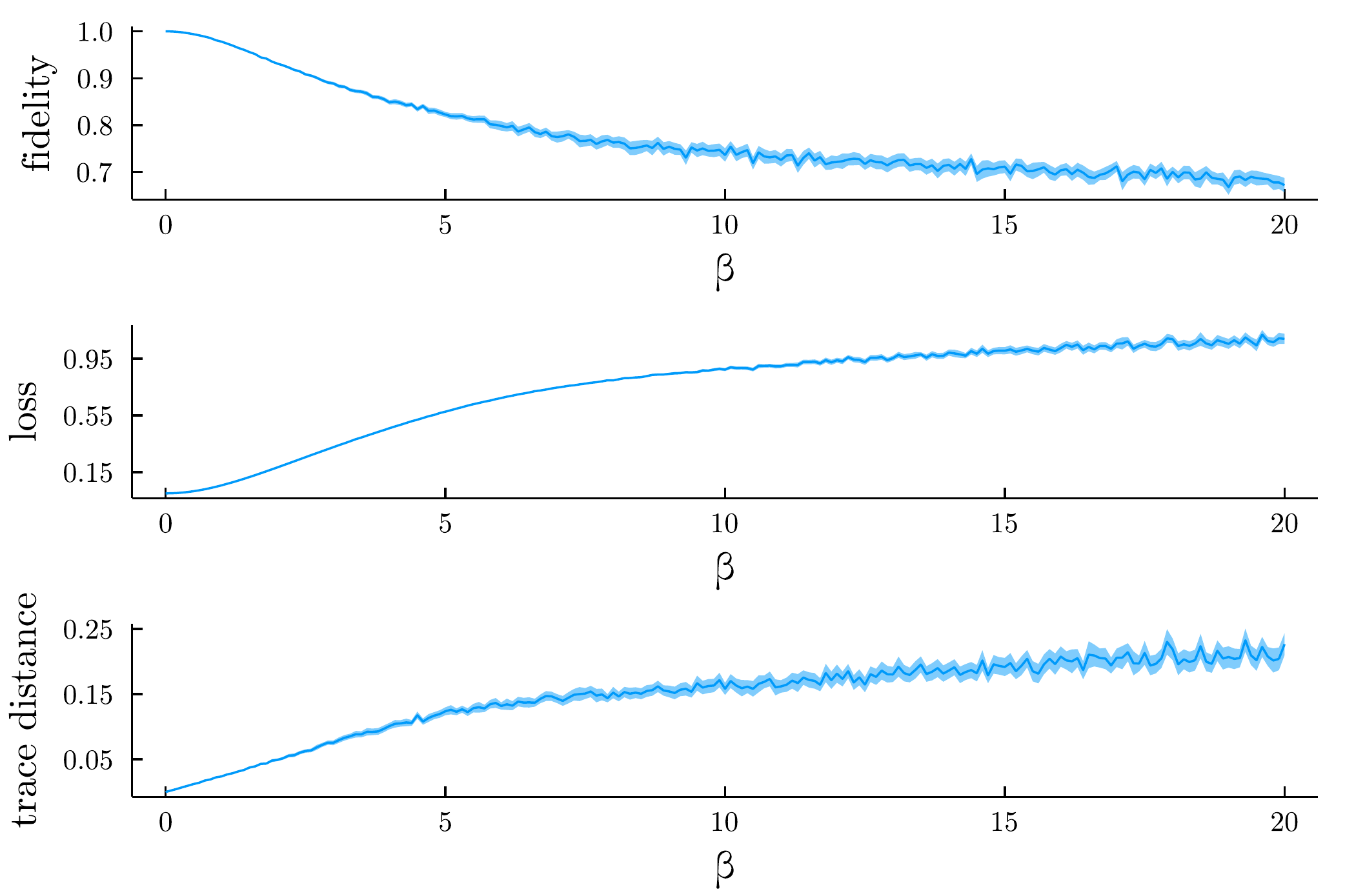}
\caption{Metrics for approximating the target thermal density $\sigma_{\beta}$ in \eqref{eq:thermal_sigma_beta} as function of $\beta$, with 500 optimization iterations using the VQT algorithm with parameters as in Fig.~\ref{fig:2d_heis_iterations}. The shaded regions indicate $95\%$ confidence intervals.}
\label{fig:2d_heis_beta}
\end{figure}

Fig.~\ref{fig:2d_heis_iterations} visualizes the convergence of the numerical method with the number of optimization steps, averaged over 50 realizations with random initial parameters $(\theta, \phi)$. One observes an unhampered, smooth convergence. The independence of the final values (after 300 iterations) of the initial random parameters indicates that the procedure is not trapped in local minima. The actual approximation metrics are shown in Fig.~\ref{fig:2d_heis_beta}, for $\beta$ between $0$ and $20$ in $0.1$ intervals, indicating that the approximation worsens with increasing $\beta$. Note that the confidence interval remains quite small throughout the experiment; thus it is likely that the particular parametric ansatz \eqref{eq:rho_ansatz_def} may not have sufficient expressibility for the two-dimensional case. Conversely, a more complex circuit or latent modular density matrix would be required for better results. This was suggested but not experimentally confirmed in the prior work \cite{Verdon2019}.

In summary, the methods introduced in this work are implemented in the Qaintum software library, which offers the functionality to construct a parametrized quantum circuit $U_{\phi}$ and apply it as conjugation \eqref{eq:rho_ansatz_def}, and then handles gradient computation internally to facilitate parameter optimization via the Flux \cite{Flux} machine learning toolbox.

\section{Conclusions and outlook}

We have demonstrated several computational advantages of the Bloch representation, in particular in the context of variational circuit and quantum channel optimization for mixed states. Nevertheless, there are cases in which a conversion between a conventional matrix representation is still required. One scenario is the task of computing the eigenvalues of a density operator, when, for example, ensuring that it is positive semidefinite. It could be possible to adapt an implementation of, say, the QR iteration algorithm, which involves conjugations as in \eqref{eq:rho_U_conjugation}, to work directly with the Bloch vector representation, but established linear algebra software packages certainly expect a matrix as input. Another scenario for a matrix representation as starting point is a ``pure state'', i.e., a density operator of the form $\rho = \ket{\psi}\bra{\psi}$ with $\ket{\psi} \in \C^{2^n}$ a state\-vector.

We remark that obtaining gradients as described in Sect.~\ref{sec:backpropagation} is computationally more efficient than the parameter shift rule \cite{Li2017, Mitarai2018, Schuld2019} in most cases; the latter is tailored to physical quantum computers, for which the intermediate quantum states are inaccessible. The parameter shift rule has the drawback that a circuit has to be run twice for each parameter. In our case, only a single backward pass through the circuit is necessary to obtain the gradients with respect to all gates. We have demonstrated the practical feasibility of this approach via the implementation of the VQT algorithm.

As an outlook, we want to draw the attention to tensor network methods as powerful tools for simulating density operators \cite{Verstraete2004, Hauschild2018}. An interesting project for future research could consist of approximating the multi-qubit Bloch vector by a real-valued matrix product state.

\acknowledgments
We thank Frank Pollmann for helpful discussions, the Munich Center for Quantum Science and Technology for support, and the Leibniz Supercomputing Centre (LRZ) for providing computing resources.

\appendix

\section{Bloch representation of common quantum logic gates and channels}
\label{sec:gates_bloch_table}

\subsection{Single-qubit gates}

Table~\ref{tab:single_qubit_gates} summarizes the Bloch representation of common single-qubit gates. Regarding the general rotation gate $R_{\vec{n}}(\theta)$, $\theta \in \R$ is the rotation angle, $\vec{n} \in \R^3$ is the unit vector specifying the rotation axis, $\vec{\sigma} = (X, Y, Z)$ the Pauli vector and $\text{Rot}(\vec{n}, \theta)$ the matrix describing a classical three-dimensional rotation by angle $\theta$. Its action on a vector $\vec{v} \in \R^3$ is given by Rodrigues' rotation formula:
\begin{multline}
\text{Rot}(\vec{n}, \theta) \vec{v} = \cos(\theta) \vec{v} + \sin(\theta) \, \vec{n} \times \vec{v} \\ + (1 - \cos(\theta)) (\vec{n} \cdot \vec{v}) \, \vec{n}.
\end{multline}
See, e.g., \cite{NielsenChuang} for a derivation that $R_{\vec{n}}(\theta)$ indeed translates to a classical rotation in the Bloch representation.

The projector $\ket{1}\bra{1}$ is not actually a unitary gate, but we include it here as ingredient of controlled gates.

\begin{table*}[p]
\begin{tabular}{ccccc}
symbol & $U$ & $\mathsf{U}$ & Bloch repr.\ of $\mathcal{S}_U$ & Bloch repr.\ of $\mathcal{A}_U$ \\
\hline
$X$ & $\begin{pmatrix} 0 & 1 \\ 1 & 0 \end{pmatrix}$ & $\begin{pmatrix} 1 & & & \\ & 1 & & \\ & & -1 & \\ & & & -1 \end{pmatrix}$ & $\begin{pmatrix} & 1 & & \\ 1 & & & \\ & & 0 & \\ & & & 0 \end{pmatrix}$ & $\begin{pmatrix} 0 & & & \\ & 0 & & \\ & & & 1 \\ & & -1 & \end{pmatrix}$ \\
$Y$ & $\begin{pmatrix} 0 & -i \\ i & 0 \end{pmatrix}$ & $\begin{pmatrix} 1 & & & \\ & -1 & & \\ & & 1 & \\ & & & -1 \end{pmatrix}$ & $\begin{pmatrix} & & 1 & \\ & 0 & & \\ 1 & & & \\ & & & 0 \end{pmatrix}$ & $\begin{pmatrix} 0 & & & \\ & & & -1 \\ & & 0 & \\ & 1 & & \end{pmatrix}$ \\
$Z$ & $\begin{pmatrix} 1 & 0 \\ 0 & -1 \end{pmatrix}$ & $\begin{pmatrix} 1 & & & \\ & -1 & & \\ & & -1 & \\ & & & 1 \end{pmatrix}$ & $\begin{pmatrix} & & & 1 \\ & 0 & & \\ & & 0 & \\ 1 & & & \end{pmatrix}$ & $\begin{pmatrix} 0 & & & \\ & & 1 & \\ & -1 & & \\ & & & 0 \end{pmatrix}$ \\
$H$ & $\frac{1}{\sqrt{2}} \begin{pmatrix} 1 & 1 \\ 1 & -1 \end{pmatrix}$ & $\begin{pmatrix} 1 & & & \\ & & & 1 \\ & & -1 & \\ & 1 & & \end{pmatrix}$ & $\begin{pmatrix} & \frac{1}{\sqrt{2}} & & \frac{1}{\sqrt{2}} \\ \frac{1}{\sqrt{2}} & & & \\ & & 0 & \\ \frac{1}{\sqrt{2}} & & & \end{pmatrix}$ & $\begin{pmatrix} 0 & & & \\ & & \frac{1}{\sqrt{2}} & \\ & -\frac{1}{\sqrt{2}} & & \frac{1}{\sqrt{2}} \\ & & -\frac{1}{\sqrt{2}} & \end{pmatrix}$ \\
$S$ & $\begin{pmatrix} 1 & 0 \\ 0 & i \end{pmatrix}$ & $\begin{pmatrix} 1 & & & \\ & & -1 & \\ & 1 & & \\ & & & 1 \end{pmatrix}$ & $\frac{1}{2} \begin{pmatrix} 1 & & & 1 \\ & 1 & -1 & \\ & 1 & 1 & \\ 1 & & & 1 \end{pmatrix}$ & $\frac{1}{2} \begin{pmatrix} -1 & & & 1 \\ & -1 & 1 & \\ & -1 & -1 & \\ 1 & & & -1 \end{pmatrix}$ \\
phase shift & $\begin{pmatrix} 1 & 0 \\ 0 & \e^{i \varphi} \end{pmatrix}$ & $\begin{pmatrix} 1 & & & \\ & \cos(\varphi) & -\sin(\varphi) & \\ & \sin(\varphi) & \cos(\varphi) & \\ & & & 1 \end{pmatrix}$ & $\begin{pmatrix} \mathfrak{c}^2 & & & \mathfrak{s}^2 \\ & \mathfrak{c}^2 & -\mathfrak{c} \mathfrak{s} & \\ & \mathfrak{c} \mathfrak{s} & \mathfrak{c}^2 & \\ \mathfrak{s}^2 & & & \mathfrak{c}^2 \end{pmatrix}$ & $\begin{pmatrix} -\mathfrak{c} \mathfrak{s} & & & \mathfrak{c} \mathfrak{s} \\ & -\mathfrak{c} \mathfrak{s} & \mathfrak{s}^2 & \\ & -\mathfrak{s}^2 & -\mathfrak{c} \mathfrak{s} & \\ \mathfrak{c} \mathfrak{s} & & & -\mathfrak{c} \mathfrak{s} \end{pmatrix}$ \\
$R_{\text{x}}(\theta)$ & $\e^{-i \theta X/2}$ & $\begin{pmatrix} 1 & & & \\ & 1 & & \\ & & \cos(\theta) & -\sin(\theta) \\ & & \sin(\theta) & \cos(\theta) \end{pmatrix}$ & $\begin{pmatrix} \mathfrak{c} & & & \\ & \mathfrak{c} & & \\ & & \mathfrak{c} & -\mathfrak{s} \\ & & \mathfrak{s} & \mathfrak{c} \end{pmatrix}$ & $\begin{pmatrix} & \mathfrak{s} & & \\ \mathfrak{s} & & & \\ & & 0 & \\ & & & 0 \end{pmatrix}$ \\
$R_{\text{y}}(\theta)$ & $\e^{-i \theta Y/2}$ & $\begin{pmatrix} 1 & & & \\ & \cos(\theta) & & \sin(\theta) \\ & & 1 & \\ & -\sin(\theta) & & \cos(\theta) \end{pmatrix}$ & $\begin{pmatrix} \mathfrak{c} & & & \\ & \mathfrak{c} & & \mathfrak{s} \\ & & \mathfrak{c} & \\ & -\mathfrak{s} & & \mathfrak{c} \end{pmatrix}$ & $\begin{pmatrix} & & \mathfrak{s} & \\ & 0 & & \\ \mathfrak{s} & & & \\ & & & 0 \end{pmatrix}$ \\
$R_{\text{z}}(\theta)$ & $\e^{-i \theta Z/2}$ & $\begin{pmatrix} 1 & & & \\ & \cos(\theta) & -\sin(\theta) & \\ & \sin(\theta) & \cos(\theta) & \\ & & & 1 \end{pmatrix}$ & $\begin{pmatrix} \mathfrak{c} & & & \\ & \mathfrak{c} & -\mathfrak{s} & \\ & \mathfrak{s} & \mathfrak{c} & \\ & & & \mathfrak{c} \end{pmatrix}$ & $\begin{pmatrix} & & & \mathfrak{s} \\ & 0 & & \\ & & 0 & \\ \mathfrak{s} & & & \end{pmatrix}$ \\
$R_{\vec{n}}(\theta)$ & $\e^{-i \theta (\vec{n} \cdot \vec{\sigma})/2}$ & $\begin{pmatrix} 1 & \\ & \text{Rot}(\vec{n}, \theta) \vphantom{\Bigg\vert} \end{pmatrix}$ & $\begin{pmatrix} \mathfrak{c} & & & \\ & \mathfrak{c} & -\mathfrak{s} n_3 & \mathfrak{s} n_2 \\ & \mathfrak{s} n_3 & \mathfrak{c} & -\mathfrak{s} n_1 \\ & -\mathfrak{s} n_2 & \mathfrak{s} n_1 & \mathfrak{c} \end{pmatrix}$ & $\begin{pmatrix} & \mathfrak{s} n_1 & \mathfrak{s} n_2 & \mathfrak{s} n_3 \\ \mathfrak{s} n_1 & & & \\ \mathfrak{s} n_2 & & & \\ \mathfrak{s} n_3 & & & \end{pmatrix}$ \\
$\ket{1}\bra{1}$ & $\begin{pmatrix} 0 & 0 \\ 0 & 1 \end{pmatrix}$ & $\frac{1}{2} \begin{pmatrix} 1 & & & -1 \\ & 0 & & \\ & & 0 & \\ -1 & & & 1 \end{pmatrix}$ & $\frac{1}{2} \begin{pmatrix} 1 & & & -1 \\ & 1 & & \\ & & 1 & \\ -1 & & & 1 \end{pmatrix}$ & $\frac{1}{2} \begin{pmatrix} 0 & & & \\ & & -1 & \\ & 1 & & \\ & & & 0 \end{pmatrix}$
\end{tabular}
\caption{Bloch representation of common single-qubit quantum gates and associated operators in \eqref{eq:SA_def}, using shorthands $\mathfrak{c} = \cos(\theta/2)$ and $\mathfrak{s} = \sin(\theta/2)$ for the rotation gates, and likewise $\mathfrak{c} = \cos(\varphi/2)$ and $\mathfrak{s} = \sin(\varphi/2)$ for the phase shift gate.}
\label{tab:single_qubit_gates}
\end{table*}

\subsection{Two-qubit gates}

Table~\ref{tab:two_qubit_rotation_gates} shows the Bloch representation of selected two-qubit gates. For conciseness of notation, we use brak-ket notation as $\ket{j k} = e_j \otimes e_k$ for $j, k = 0, \dots, 3$, with $e_j$ the $j$-th unit vector of length $4$.

\begin{table*}[p]
\begin{tabular}{c@{\hskip 0.35cm}c@{\hskip 0.35cm}cp{7cm}}
symbol & $U$ & $\mathsf{U}$ \\
\hline
$R_{\text{xx}}(\theta)$ & $\e^{-i \theta X \otimes X / 2}$ &
$\begin{pmatrix}
 1 &   &   &   &   &   &   &   &   &   &   &   &   &   &   &   \\
   & 1 &   &   &   &   &   &   &   &   &   &   &   &   &   &   \\
   &   & c &   &   &   &   &-s &   &   &   &   &   &   &   &   \\
   &   &   & c &   &   & s &   &   &   &   &   &   &   &   &   \\
   &   &   &   & 1 &   &   &   &   &   &   &   &   &   &   &   \\
   &   &   &   &   & 1 &   &   &   &   &   &   &   &   &   &   \\
   &   &   &-s &   &   & c &   &   &   &   &   &   &   &   &   \\
   &   & s &   &   &   &   & c &   &   &   &   &   &   &   &   \\
   &   &   &   &   &   &   &   & c &   &   &   &   &-s &   &   \\
   &   &   &   &   &   &   &   &   & c &   &   &-s &   &   &   \\
   &   &   &   &   &   &   &   &   &   & 1 &   &   &   &   &   \\
   &   &   &   &   &   &   &   &   &   &   & 1 &   &   &   &   \\
   &   &   &   &   &   &   &   &   & s &   &   & c &   &   &   \\
   &   &   &   &   &   &   &   & s &   &   &   &   & c &   &   \\
   &   &   &   &   &   &   &   &   &   &   &   &   &   & 1 &   \\
   &   &   &   &   &   &   &   &   &   &   &   &   &   &   & 1 \\
\end{pmatrix}$ &
$= I_{16} \newline %
- (1 - c) \big( \ket{0 2}\bra{0 2} + \ket{2 0}\bra{2 0} + \ket{0 3}\bra{0 3} + \ket{3 0}\bra{3 0} + \ket{1 2}\bra{1 2} + \ket{2 1}\bra{2 1} + \ket{1 3}\bra{1 3} + \ket{3 1}\bra{3 1} \big) \newline %
+ s \big( \ket{0 3} \bra{1 2} - \ket{1 2}\bra{0 3} + \ket{3 0} \bra{2 1} - \ket{2 1}\bra{3 0} + \ket{1 3}\bra{0 2} - \ket{0 2}\bra{1 3} + \ket{3 1}\bra{2 0} - \ket{2 0}\bra{3 1} \big)$ \\
\vspace{0.1cm} \\
$R_{\text{yy}}(\theta)$ & $\e^{-i \theta Y \otimes Y / 2}$ &
$\begin{pmatrix}
 1 &   &   &   &   &   &   &   &   &   &   &   &   &   &   &   \\
   & c &   &   &   &   &   &   &   &   &   & s &   &   &   &   \\
   &   & 1 &   &   &   &   &   &   &   &   &   &   &   &   &   \\
   &   &   & c &   &   &   &   &   &-s &   &   &   &   &   &   \\
   &   &   &   & c &   &   &   &   &   &   &   &   &   & s &   \\
   &   &   &   &   & 1 &   &   &   &   &   &   &   &   &   &   \\
   &   &   &   &   &   & c &   &   &   &   &   & s &   &   &   \\
   &   &   &   &   &   &   & 1 &   &   &   &   &   &   &   &   \\
   &   &   &   &   &   &   &   & 1 &   &   &   &   &   &   &   \\
   &   &   & s &   &   &   &   &   & c &   &   &   &   &   &   \\
   &   &   &   &   &   &   &   &   &   & 1 &   &   &   &   &   \\
   &-s &   &   &   &   &   &   &   &   &   & c &   &   &   &   \\
   &   &   &   &   &   &-s &   &   &   &   &   & c &   &   &   \\
   &   &   &   &   &   &   &   &   &   &   &   &   & 1 &   &   \\
   &   &   &   &-s &   &   &   &   &   &   &   &   &   & c &   \\
   &   &   &   &   &   &   &   &   &   &   &   &   &   &   & 1 \\
\end{pmatrix}$ &
$= I_{16} \newline %
- (1 - c) \big( \ket{0 1}\bra{0 1} + \ket{1 0}\bra{1 0} + \ket{0 3}\bra{0 3} + \ket{3 0}\bra{3 0} + \ket{1 2}\bra{1 2} + \ket{2 1}\bra{2 1} + \ket{2 3}\bra{2 3} + \ket{3 2}\bra{3 2}\big) \newline %
+ s \big( \ket{0 1}\bra{2 3} - \ket{2 3}\bra{0 1} + \ket{1 0}\bra{3 2} - \ket{3 2}\bra{1 0} + \ket{1 2}\bra{3 0} - \ket{3 0}\bra{1 2} + \ket{2 1}\bra{0 3} - \ket{0 3}\bra{2 1} \big)$ \\
\vspace{0.1cm} \\
$R_{\text{zz}}(\theta)$ & $\e^{-i \theta Z \otimes Z / 2}$ &
$\begin{pmatrix}
 1 &   &   &   &   &   &   &   &   &   &   &   &   &   &   &   \\
   & c &   &   &   &   &   &   &   &   &   &   &   &   &-s &   \\
   &   & c &   &   &   &   &   &   &   &   &   &   & s &   &   \\
   &   &   & 1 &   &   &   &   &   &   &   &   &   &   &   &   \\
   &   &   &   & c &   &   &   &   &   &   &-s &   &   &   &   \\
   &   &   &   &   & 1 &   &   &   &   &   &   &   &   &   &   \\
   &   &   &   &   &   & 1 &   &   &   &   &   &   &   &   &   \\
   &   &   &   &   &   &   & c &-s &   &   &   &   &   &   &   \\
   &   &   &   &   &   &   & s & c &   &   &   &   &   &   &   \\
   &   &   &   &   &   &   &   &   & 1 &   &   &   &   &   &   \\
   &   &   &   &   &   &   &   &   &   & 1 &   &   &   &   &   \\
   &   &   &   & s &   &   &   &   &   &   & c &   &   &   &   \\
   &   &   &   &   &   &   &   &   &   &   &   & 1 &   &   &   \\
   &   &-s &   &   &   &   &   &   &   &   &   &   & c &   &   \\
   & s &   &   &   &   &   &   &   &   &   &   &   &   & c &   \\
   &   &   &   &   &   &   &   &   &   &   &   &   &   &   & 1 \\
\end{pmatrix}$ &
$= I_{16} \newline %
- (1 - c) \big( \ket{0 1}\bra{0 1} + \ket{1 0}\bra{1 0} + \ket{0 2}\bra{0 2} + \ket{2 0}\bra{2 0} + \ket{1 3}\bra{1 3} + \ket{3 1}\bra{3 1} + \ket{2 3}\bra{2 3} + \ket{3 2}\bra{3 2} \big) \newline %
+ s \big( \ket{0 2}\bra{3 1} - \ket{3 1}\bra{0 2} + \ket{2 0}\bra{1 3} - \ket{1 3}\bra{2 0} + \ket{2 3}\bra{1 0} - \ket{1 0}\bra{2 3} + \ket{3 2}\bra{0 1} - \ket{0 1}\bra{3 2} \big)$
\end{tabular}
\caption{Bloch representation of selected two-qubit quantum gates, using the shorthand notations $c = \cos(\theta)$ and $s = \sin(\theta)$.}
\label{tab:two_qubit_rotation_gates}
\end{table*}

\subsection{Single-qubit quantum channels}

Table~\ref{tab:single_qubit_channels} summarizes the Bloch representation of several single-qubit quantum channels \cite{NielsenChuang}; the parameters $p, \gamma, \lambda$ are from the interval $[0, 1]$, and can be interpreted as probabilities.
\begin{table}
\begin{tabular}{p{1.5cm}p{3cm}c}
channel & Kraus operators & Bloch repr.~\eqref{eq:bloch_repr_channel} \\
\hline
bit flip & $E_0 = \sqrt{p} I_2$, \newline $E_1 = \sqrt{1 - p} X$ & $\begin{pmatrix} 1 & & & \\ & 1 & & \\ & & 2 p - 1 & \\ & & & 2 p - 1 \end{pmatrix}$ \\
phase flip & $E_0 = \sqrt{p} I_2$, \newline $E_1 = \sqrt{1 - p} Z$ & $\begin{pmatrix} 1 & & & \\ & 2 p - 1 & & \\ & & 2 p - 1 & \\ & & & 1 \end{pmatrix}$ \\
depolar\-izing channel & $E_0 = \sqrt{1 - 3p/4} I_2$, \newline $E_1 = \sqrt{p} X / 2$, \newline $E_2 = \sqrt{p} Y / 2$, \newline $E_3 = \sqrt{p} Z / 2$ & $\begin{pmatrix} 1 & & & \\ & 1 - p & & \\ & & 1 - p & \\ & & & 1 - p \end{pmatrix}$ \\
amplitude \newline damping & $E_0 = \begin{pmatrix} 1 & 0 \\ 0 & \sqrt{1 - \gamma} \end{pmatrix}$, \newline $E_1 = \begin{pmatrix} 0 & \sqrt{\gamma} \\ 0 & 0 \end{pmatrix}$ & $\begin{pmatrix} 1 & & & \\ & \sqrt{1 - \gamma} & & \\
 & & \sqrt{1 - \gamma} & \\ \gamma & & & 1 - \gamma \end{pmatrix}$ \\
phase damping & $E_0 = \begin{pmatrix} 1 & 0 \\ 0 & \sqrt{1 - \lambda} \end{pmatrix}$, \newline $E_1 = \begin{pmatrix} 0 & 0 \\ 0 & \sqrt{\lambda} \end{pmatrix}$ & $\begin{pmatrix} 1 & & & \\ & \sqrt{1 - \lambda} & & \\ & & \sqrt{1 - \lambda} & \\ & & & 1 \end{pmatrix}$ \\
\end{tabular}
\caption{Bloch representation of several single-qubit quantum channels describing noise processes.}
\label{tab:single_qubit_channels}
\end{table}

\section{Expansion of $\mathcal{S}_{F \otimes G}$ and $\mathcal{A}_{F \otimes G}$ and generalization to multiple tensor products}
\label{sec:SA_expansion}

We first verify the relations in Eqs.~\eqref{eq:SA_expansion}. Given complex matrices $F \in \C^{m \times m}$ and $G \in \C^{n \times n}$, note that the linear operators $\mathcal{S}_{F \otimes G}$ and $\mathcal{A}_{F \otimes G}$ act on Hermitian matrices $\rho$ of dimension $m n \times m n$. For any such $\rho$, one calculates
\begin{equation}
\begin{split}
&(\mathcal{S}_F \otimes \mathcal{S}_G)(\rho) - (\mathcal{A}_F \otimes \mathcal{A}_G)(\rho) \\
&= \frac{1}{2} \bigg( F \otimes I \cdot \frac{1}{2}(I \otimes G \cdot \rho + \rho \cdot I \otimes G^{\dagger}) \\
& \quad + \frac{1}{2}(I \otimes G \cdot \rho + \rho \cdot I \otimes G^{\dagger}) \cdot F^{\dagger} \otimes I \bigg) \\
&- \frac{i}{2} \bigg( F \otimes I \cdot \frac{i}{2}(I \otimes G \cdot \rho - \rho \cdot I \otimes G^{\dagger}) \\
& \quad - \frac{i}{2}(I \otimes G \cdot \rho - \rho \cdot I \otimes G^{\dagger}) \cdot F^{\dagger} \otimes I \bigg) \\
&= \frac{1}{2} (F \otimes G \cdot \rho + \rho \cdot F^{\dagger} \otimes G^{\dagger})
 = \mathcal{S}_{F \otimes G}(\rho)
\end{split}
\end{equation}
and
\begin{equation}
\begin{split}
&(\mathcal{S}_F \otimes \mathcal{A}_G)(\rho) + (\mathcal{A}_F \otimes \mathcal{S}_G)(\rho) \\
&= \frac{1}{2} \bigg( F \otimes I \cdot \frac{i}{2}(I \otimes G \cdot \rho - \rho \cdot I \otimes G^{\dagger}) \\
& \quad + \frac{i}{2}(I \otimes G \cdot \rho - \rho \cdot I \otimes G^{\dagger}) \cdot F^{\dagger} \otimes I \bigg) \\
&+ \frac{i}{2} \bigg( F \otimes I \cdot \frac{1}{2}(I \otimes G \cdot \rho + \rho \cdot I \otimes G^{\dagger}) \\
& \quad - \frac{1}{2}(I \otimes G \cdot \rho + \rho \cdot I \otimes G^{\dagger}) \cdot F^{\dagger} \otimes I \bigg) \\
&= \frac{i}{2}(F \otimes G \cdot \rho - \rho \cdot F^{\dagger} \otimes G^{\dagger}) = \mathcal{A}_{F \otimes G}(\rho).
\end{split}
\end{equation}

Recursive application of \eqref{eq:SA_expansion} facilitates a generalization to multiple tensor products, i.e., an expansion of $\mathcal{S}_{G_{n-1} \otimes \cdots \otimes G_0}$ and $\mathcal{A}_{G_{n-1} \otimes \cdots \otimes G_0}$ for complex matrices $G_0, \dots, G_{n-1}$. To arrive at a concise expression, observe that \eqref{eq:SA_expansion} formally resembles the product of two complex numbers, with $\mathcal{S}$ and $\mathcal{A}$ playing the roles of the real and imaginary parts, respectively. Following this analogy, let $z_j = x_j + i y_j$ with $x_j, y_j \in \R$ for $j = 0, \dots, n-1$. Then the product of the $z_j$'s in terms of real and imaginary parts is
\begin{equation}
z_{n-1} \cdots z_1 z_0 = \begin{pmatrix} x_{n-1} & -y_{n-1} \\ y_{n-1} & x_{n-1} \end{pmatrix} \cdots \begin{pmatrix} x_1 & -y_1 \\ y_1 & x_1 \end{pmatrix} \begin{pmatrix} x_0 \\ y_0 \end{pmatrix}
\end{equation}
when identifying $\C \simeq \R^2$. Thus likewise
\begin{multline}
\label{eq:SA_mpo_expansion}
\begin{pmatrix} \mathcal{S}_{G_{n-1} \otimes \cdots \otimes G_0} \\ \mathcal{A}_{G_{n-1} \otimes \cdots \otimes G_0} \end{pmatrix} \\
= \begin{pmatrix} \mathcal{S}_{G_{n-1}} & -\mathcal{A}_{G_{n-1}} \\ \mathcal{A}_{G_{n-1}} & \mathcal{S}_{G_{n-1}} \end{pmatrix} \cdots \begin{pmatrix} \mathcal{S}_{G_1} & -\mathcal{A}_{G_1} \\ \mathcal{A}_{G_1} & \mathcal{S}_{G_1} \end{pmatrix} \begin{pmatrix} \mathcal{S}_{G_0} \\ \mathcal{A}_{G_0} \end{pmatrix},
\end{multline}
with $\mathcal{S}_{G_j}$ and $\mathcal{A}_{G_j}$ understood to act on the $j$-th qubit. We remark that \eqref{eq:SA_mpo_expansion} is in fact a matrix product operator representation with virtual bond dimension $2$.

\section{Benchmarking}
\label{sec:benchmarking}

\begin{figure}[!ht]
\centering
\subfloat[Pauli gates]{\includegraphics[width=\columnwidth]{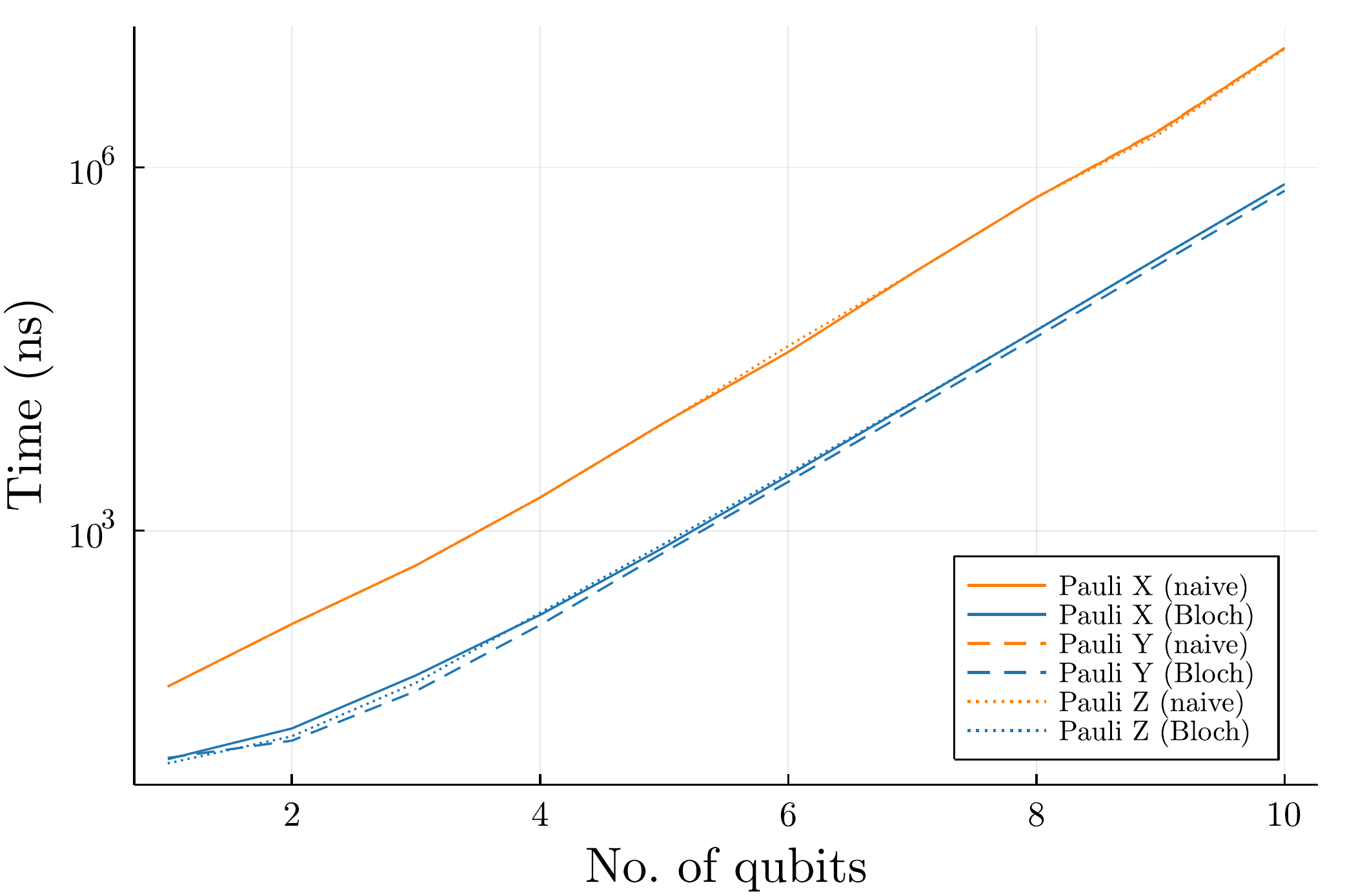}} \\
\subfloat[rotation operators]{\includegraphics[width=\columnwidth]{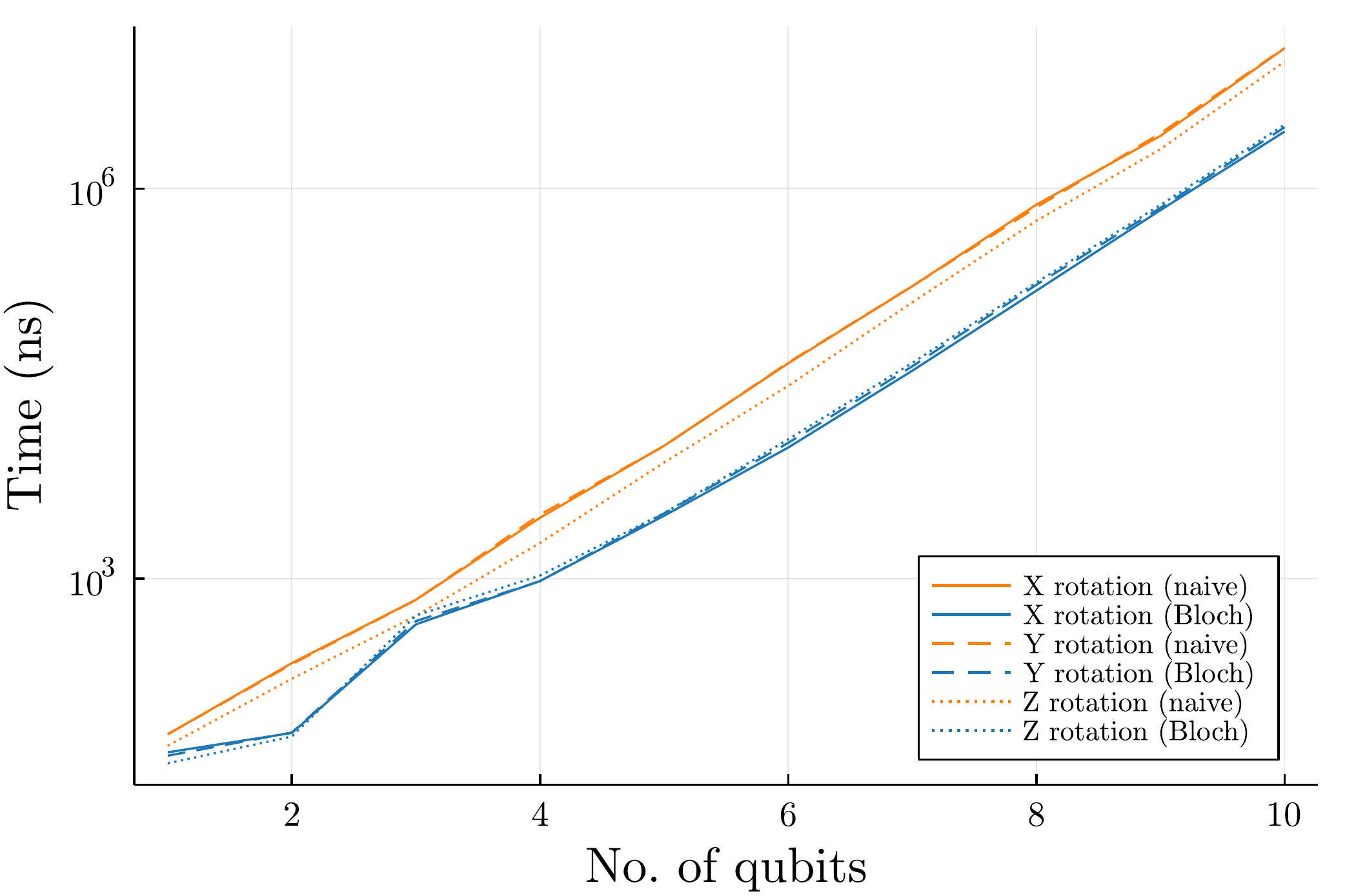}} \\
\subfloat[two-qubit gates]{\includegraphics[width=\columnwidth]{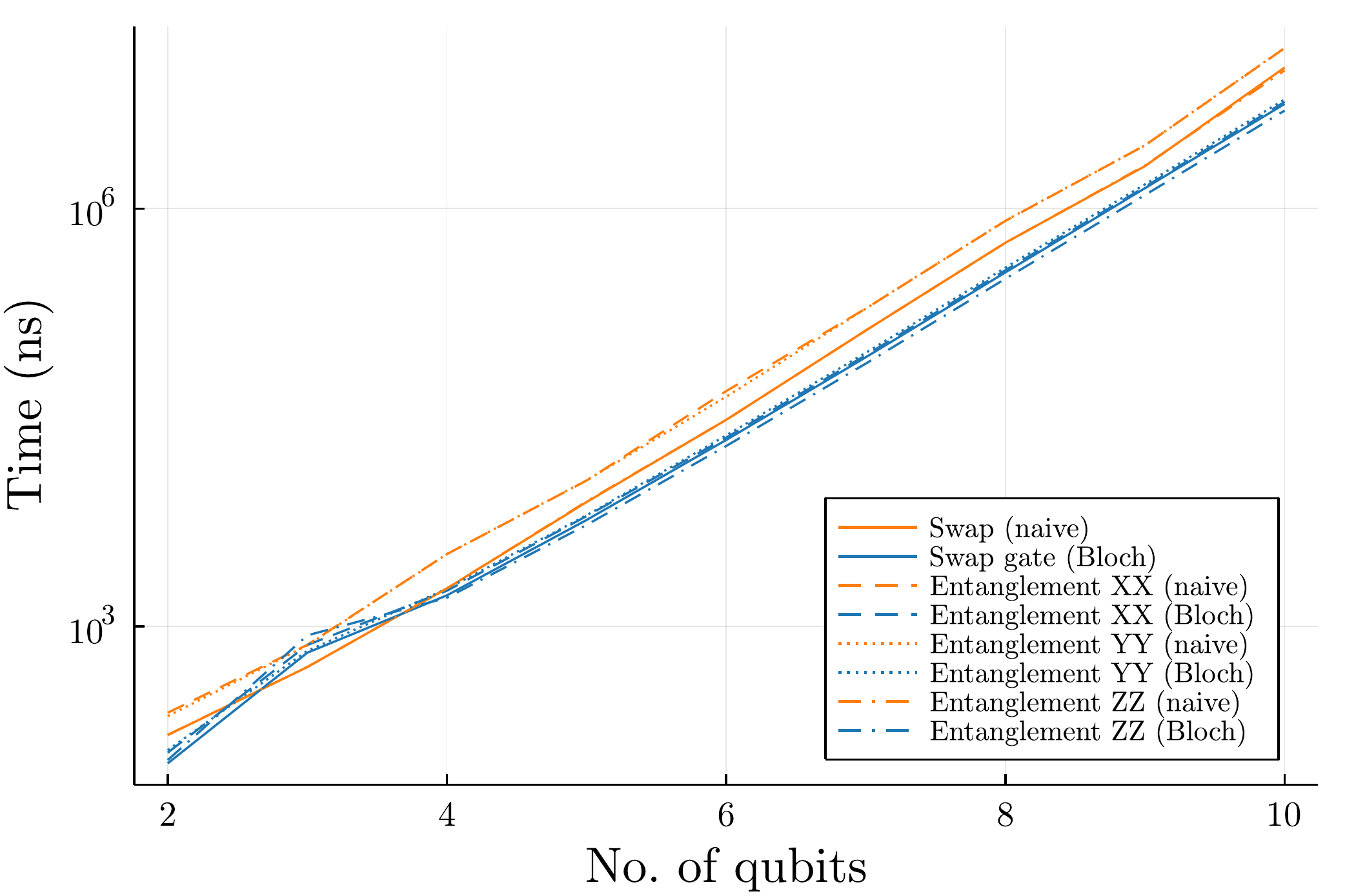}}
\caption{Benchmark comparison of our Bloch vector methodology (blue) with a conventional implementation by matrix conjugations (orange), for the task of applying typical single- and two-qubit gates.}
\label{fig:benchmarks}
\end{figure}

We compare the runtime of our Bloch representation for applying unitary quantum gates with a conventional conjugation of the density matrix, see Eq.~\eqref{eq:rho_U_conjugation}. Due to the highly sparse nature of the gates depicted in Appendix~\ref{sec:gates_bloch_table}, we are able to implement Eq.~\eqref{eq:bloch_repr_unitary} in a matrix-free manner using a single loop over the stored Bloch vector, which potentially offers a $\mathcal{O}(1)$ speedup. (The asymptotic computational complexity is linear in the number of Bloch vector entries for both versions.) A comparison of the application of single and two-qubit gates using our methodology, and an optimized in-place multiplication code using Julia's SparseArrays module (sparse CSC format for the gates and dense format for the density matrix) is shown in Fig.~\ref{fig:benchmarks}. Indeed one observes a constant speedup facilitated by the Bloch representation. We note some possible cache optimization issues for the case of three qubits, which could be remedied by proper chunking of the stored Bloch vectors. Single-qubit gates exhibit the largest runtime advantage, while the smaller speedup for controlled gates is likely due to the more involved expansion of such gates, see Sect.~\ref{sec:controlled_gates}.

The benchmarking was performed on the cloud computing nodes offered by the Leibniz Supercomputing Centre; specifically for this study single-threaded on a Intel(R) Xeon(R) Gold 6148 CPU @ 2.40GHz.

\end{document}